\documentclass[nofootinbib,twocolumn,aps,pre,superscriptaddress,citeautoscript,floatfix, groupedaddress]{revtex4-1}

\usepackage{graphicx}
\usepackage{amsmath,amssymb}
\usepackage[dvipsnames]{xcolor}
\usepackage{float}
\usepackage{array}

\begin{document}
\title{Critical behavior of charge-regulated macro-ions}
\author{Yael Avni $^{1}$}
\author{Rudolf Podgornik$^{2,3,4}$}
\author{David Andelman$^{1,}\footnote{Corresponding author; ~e-mail: andelman@tau.ac.il}$}
\affiliation{${}^{1}$Raymond and Beverly Sackler School of Physics and Astronomy,\\ Tel Aviv University, Ramat Aviv 69978, Tel Aviv, Israel}
\affiliation{${}^{2}$School of Physical Sciences and Kavli Institute for Theoretical Sciences, University of Chinese Academy of Sciences, Beijing 100049, China}
\affiliation{${}^{3}$CAS Key Laboratory of Soft Matter Physics, Institute of Physics, Chinese Academy of Sciences (CAS), Beijing 100190, China}
\affiliation{${}^{4}$Department of Physics, Faculty of Mathematics and Physics, University of Ljubljana, SI-1000 Ljubljana, Slovenia}
%%%%%%%%%%%%%%%%%%%%%%%%%%

\begin{abstract}
%%%%%%%%%%%%%%%%%%%%%%%%%%%%%%%%%%
Based on a collective description of electrolytes composed of charge-regulated macro-ions and simple salt ions, we analyze their equilibrium charge state in the bulk and their behavior in the vicinity of an external electrified surface. The mean-field formulation of mobile macro-ions in an electrolyte bathing solution is extended to include interactions between association/dissociation sites. We demonstrate that above a critical concentration of salt, and similar to the critical micelle concentration, a non-trivial distribution of charge states sets in. Such a charge state can eventually lead to a liquid-liquid phase separation based on charge regulation.
\end{abstract}
%%%%%%%%%%%%%%%%%%%%%%%%%%%%%%%%%%

\maketitle

%%%%%%%%%
\section{Introduction}
%%%%%%%%%
The Derjaguin-Landau-Verwey-Overbeek (DLVO) theory identifies the interplay between attractive Lifshitz - van der Waals (vdW) fluctuation forces (based on electrodynamics) and repulsive Poisson-Boltzmann  (PB) electrostatic forces as the crucial ingredients controlling the stability of colloidal suspensions in aqueous electrolyte solutions~\cite{Tomer2020}. As much as DLVO is recognized as a seminal work, it was shown later on to exhibit shortcomings on the level of its formal methodology and its model assumptions~\cite{Naji-perspective}. However, in spite those shortcomings, it has been noted more recently by Borkovec and collaborators~\cite{Trefalt2017,Trefalt2020} that the classical DLVO theory still provides a surprisingly accurate description of the measured colloidal interactions when complemented with numerical solutions of the PB equation and with {\it fully implemented charge regulation}.

Charge regulation (CR) refers to ion exchange between dissociable macromolecular moieties and their bathing electrolyte solution~\cite{Ninham1970, Koper2001,Avni2019,Bakhshandeh2019}.  The CR mechanism can be derived either via the law of mass action \cite{Trefalt2015}, or by modifying the free-energy that describes the dissociation process \cite{Podgornik1995}, in order to include any non-electrostatic interactions. The latter approach is particularly well suited to formulate a generalized PB theory for interacting CR macromolecular surfaces \cite{Podgornik1995,Harries2006}, and extending in this way the CR phenomenology. On the other hand, the theory of interacting {\it mobile} CR macro-ions in an electrolyte solution was analyzed in less detail. In several works, it was done on the level of the {\it cell model}~\cite{CellModel1, CellModel2, Koper2001}, which treats the macro-ions as fixed and is less appropriate to describe collective phenomena. More recently, a general mean-field (MF) formalism based on a collective description was put forward, accounting for the effects of mobile CR macro-ions in dilute solutions~\cite{Avni2018, Markovich2018}. The macro-ions are treated consistently as mobile point-like particles with full translational entropy, while still retaining their relevant internal degrees of freedom responsible for the CR processes. In this approach, CR is easily extended to include, for example, the pH-dependent protonation/deprotonation mechanism, relevant to protein electrostatics. Through this mechanism, the proteins respond to the presence of other proteins, nucleic acids, and molecular surfaces, for details see Ref.~\cite{Lund2013}.

The model of mobile CR macro-ions assumed so far~\cite{Avni2018, Markovich2018} a simple CR mechanism, where each ion association/dissociation is related to a constant free-energy gain, independent of the number of associated sites. However, interactions between different adsorption sites, stemming from more complex chemical reactions, vdW forces, conformational changes and cooperative processes, have important implications on the collective behavior of CR systems.

Hence, this observation serves as a motivation to our present work, where we specifically account for the interaction between the macro-ion association/dissociation sites. We explore the bulk behavior and the interfacial properties in presence of a charged surface ({\it i.e.}, an electrode), and in particular, we study the {\it thermodynamic charge states}. Above a critical concentration, it is shown that an additional charge state can develop. Under certain symmetry conditions, a bimodal phase can be obtained, in which the macro-ions can be either positively or negatively charged. In relation to a recent thermodynamic analysis of pH-driven phase separations~\cite{Prost2020}, we suggest that the emergence of distinct charge states could lead to a phase separation in this complex charged system.

The outline of the paper is as follows: In Section~\ref{model}, we present a general approach of treating solutions containing mobile CR macro-ions {\it via} an additional free-energy contribution. We then restrict ourselves to symmetric CR models containing interactions between different adsorption sites, and formulate the equations that govern the bulk behavior in the limit of large number of association sites. In Section~\ref{Results}, we introduce the phase space of the bulk solution, and identify a critical salt concentration, above which the system behavior undergoes an abrupt change. We finally calculate the screening length and the spatial distribution of macro-ions close to a charged interface, near the critical concentration. In Section~\ref{Conclusions}, we conclude by discussing possible connections between our model and the emergence of a liquid-liquid phase separation.

%%%%%%%%%
\section{The model} \label{model}
%%%%%%%%%%
Consider an electrolyte solution composed of simple monovalent ions and macro-ions dissolved in an aqueous solvent having a dielectric constant $\varepsilon$. The solution is in thermal equilibrium at temperature $T$. The concentration of positive ions, negative ions and macro-ions is denoted by $n_+$, $n_-$ and $P$, respectively. In the absence of external electric fields, the system is homogeneous, with bulk concentrations $n_{+}^{\rm b}$, $n_{-}^{\rm b}$ and $P_{\rm b}$.

\begin{figure}
\includegraphics[width = 1\columnwidth,draft=false]{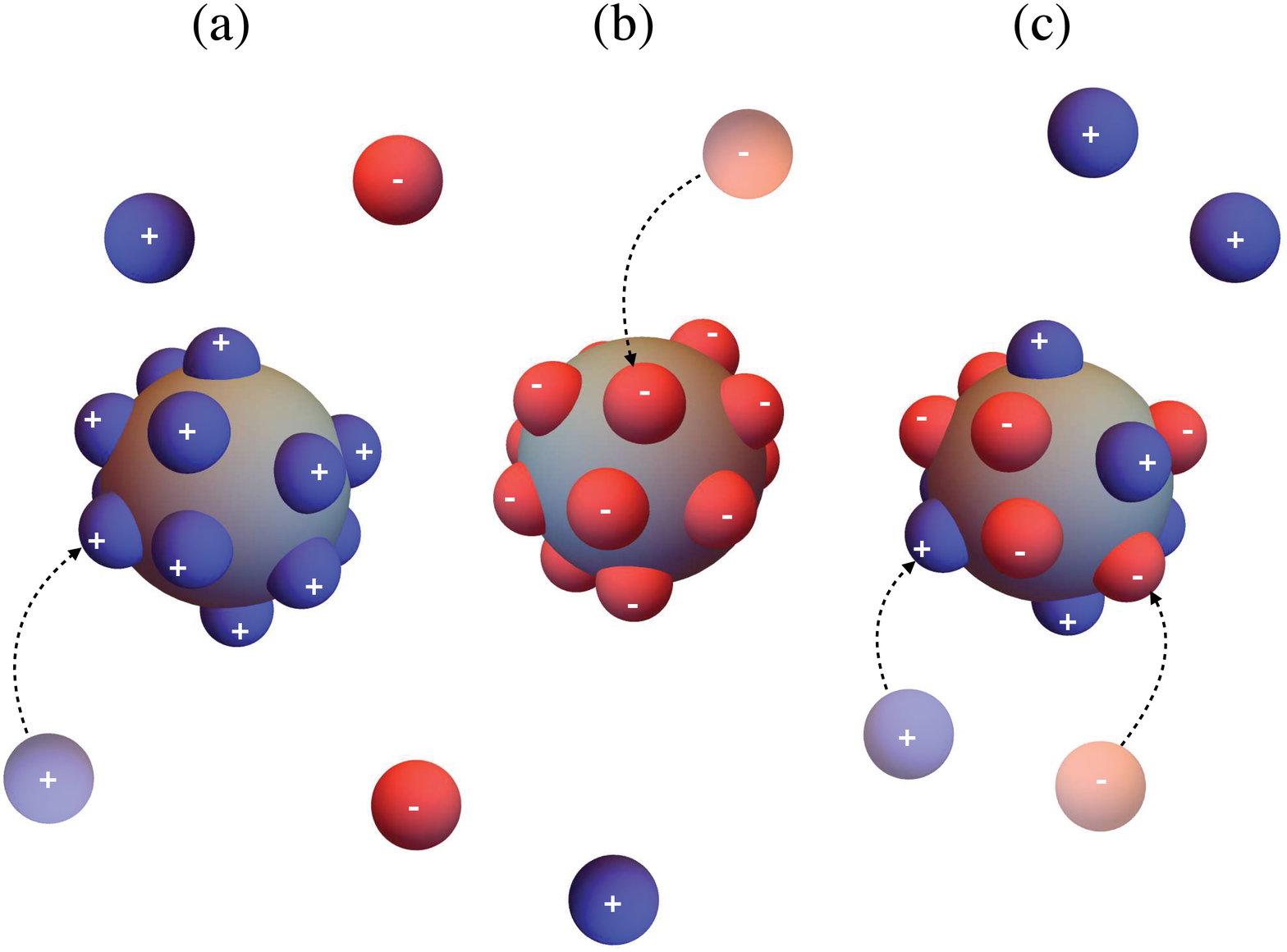} %0.5
\caption{\textsf{(Color online) Schematic drawing of three possible macro-ion charge states due to association of ions from the bathing solution: (a) a positively charged macro-ion, (b) a negatively charged macro-ion, and (c) a macro-ion with both positive and negative charges, resulting in an overall neutral macro-ion.}}
\label{Fig1}
\end{figure}

While the charge on the small ions is fixed and equals to $\pm e$, where $e$ is the elementary charge, the charge of the macro-ions can vary due to its dissociable groups, via association/dissociation of monovalent ions from/to the solution (see Fig.~\ref{Fig1}). We assume that each macro-ion has $N_+$ groups that can either be neutral, or adsorb a positive ion and become positively charged, and similarly, $N_-$ groups that can either be neutral, or adsorb a negative ion and become negatively charged. We refer to such groups as ``sites". The macro-ion can exhibit large deviations of its net charge so that {\it a priori} one does not know whether this charge is small or large. For simplicity, we identify the dissociating ions to be of the same type of the monovalent salt. This assumption simplifies the calculation, but is not an essential one, and can be easily relaxed by explicitly formulating the CR mechanism in the 
protonation/deprotonation language. In addition, we make two important simplifications for the model: (i) the macro-ions are treated as point-like; (ii) our model relies on the mean-field (MF) approximation.
These rather common simplifications make the formalism fairly straightforward to implement, but they
limit the validity of our model to dilute solutions bearing macro-ions with relatively low charge~\cite{Naji-perspective}.

 %%%%%%%%%
%Fig1
%%%%%%%%%%%%%%%%%%%%%%%%%%%%%%%%%%%%%%%%%%%
%%%%%%%%%%%%%%%%%%%%%%%%%%%%%%%%%%%%%%%%%%%

The MF free-energy, $\mathcal{F}$, can be written as~\cite{Avni2019,Tomer2020}
\begin{equation} \label{eq1}
\begin{split}
\mathcal{F} & =\int\bigg[-\frac{\varepsilon_{0}\varepsilon}{2}\left(\nabla\psi\right)^{2}+e\left(n_{+}-n_{-}\right)\psi-\mu_{+}n_{+}-\mu_{-}n_{-}\\
 & -\mu_{P}P-ST+\sum_{z_{+}=0}^{N_{+}}\sum_{z_{-}=0}^{N_{-}}p(z_{+},z_{-})g\left(z_{+},z_{-}\right)\bigg]{\rm d^3}{r}
\end{split}
\end{equation}
where $\psi(\bf{r})$ is the electrostatic potential, $\mu_{\pm}$ and $\mu_P$ incorporate the chemical potentials of the monovalent ions and the macro-ions, and $S$ is the total translational entropy. Within the aforementioned point-like approximation, $S$ is identified as the ideal-gas entropy,
\begin{equation} \label{eq2}
\begin{split}
S/k_{{\rm B}} & =-\sum_{i=\pm}n_{i}\left[\ln\left(n_{i}\lambda_i^{3}\right)-1\right]\\
 & -\sum_{z_{+}=0}^{N_{+}}\sum_{z_{-}=0}^{N_{-}}p(z_{+},z_{-})\left[\ln\left(p(z_{+},z_{-})\lambda_P^{3}\right)-1\right],
\end{split}
\end{equation}
with $k_{\rm B}$ being the Boltzmann constant and ${\lambda_{\pm}=h/\sqrt{2\pi m_{\pm} k_{\rm B}T}}$ and ${\lambda_{P}=h/\sqrt{2\pi m_{P} k_{\rm B}T}}$ the thermal de Broglie wavelength of the positive/negative ions and macro-ions, respectively, where $m_+$, $m_-$ and $m_P$ are their respective masses. We further assume that the small ions have equal masses, $m\equiv m_+=m_-$, and define ${\lambda_{+}=\lambda_{-}\equiv\lambda}$.
In Eqs.~(\ref{eq1}) and~(\ref{eq2}), $p(z_{+},z_{-})$ denotes the concentration of macro-ions with ${0\leq z_{+}\leq N_+}$ positively charged sites and ${0\leq z_{-}\leq N_-}$ negatively charged ones, as the maximal $z_+$ ($z_-$) valency is simply $N_+$ ($N_-$). The different macro-ion states appear in the entropy as different particle species due to their distinguishability, but they are related to one another by the normalizing condition
\begin{equation} \label{eq3}
\sum_{z_{+},z_{-}}p(z_{+},z_{-})=P,
\end{equation}
where the shorthand notation $\sum_{z_{+}=0}^{N_+}\sum_{z_{-}=0}^{N_-} \to \sum_{z_{+},z_{-}}$ is used.
Note that only the last term in Eq.~(\ref{eq1}) distinguishes a simple system, of macro-ions with a fixed charge, from a CR one. For each macro-ion state characterized by a specific set of $(z_+,z_-)$, this term contains the respective macro-ion concentration multiplied by the free-energy of the macro-ion internal state, $g(z_+,z_-)$.
In its most general form, the internal state free-energy
can be written as
\begin{equation} \label{eq4}
\begin{split}
g\left(z_{+},z_{-}\right)= & e\left(z_{+}-z_{-}\right)\psi-k_{{\rm B}}T\ln\left[{N_{+} \choose z_{+}}{N_{-} \choose z_{-}}\right]\\
 & +F_{\rm CR}\left(z_{+},z_{-}\right)-\mu_{+}z_{+}-\mu_{-}z_{-}.
\end{split}
\end{equation}
The first two terms in Eq.~(\ref{eq4}) are the electrostatic energy of the macro-ion and its internal entropy, accounting for different ways to arrange $z_{+}$ positively charged sites and $z_{-}$ negatively charged sites on each macro-ion. The third term, $F_{\rm CR}(z_+,z_-)$, is the free-energy gain from the association/dissociation process. This phenomenological term includes the energy gain from chemical reactions, vdW forces, conformational changes, cooperative processes and possibly others, as well as the entropy within a single site (note that the entropy of mixing between the different sites was already taken into account separately by the second term). Finally, the last two terms incorporate the chemical potentials of the adsorbed positive and negative ions.

Minimizing the free-energy with respect to $n_+$, $n_-$ and $p(z_+,z_-)$, we obtain
\begin{equation} \label{eq5}
\begin{split}
n_{\pm}({\bf r}) & =n_{\pm}^{{\rm b}}{\rm e}^{\mp\beta e\psi\left({\bf r}\right)}\\
p(z_{+},z_{-};{\bf r}) & =p_{{\rm b}}(z_{+},z_{-}){\rm e}^{-\beta\left(z_{+}-z_{-}\right)e\psi\left({\bf r}\right)},
\end{split}
\end{equation}
where $\beta=1/k_{\rm B} T$ and $p_{\rm b}(z_{+},z_{-})$ is the restricted bulk concentration of macro-ions having ${0 \leq z_+ \leq N_+}$ positive and ${0 \leq z_- \leq N_-}$ negative charges. In deriving Eq.~(\ref{eq5}), the chemical potentials acted as Lagrange multipliers, enforcing the constraint that all concentrations reach their bulk values at the bulk reference potential, $\psi=0$. More explicitly, $\mu_{\pm}$ and $\mu_{P}$ satisfy the equations
\begin{equation} \label{eq6}
\begin{split}
\mu_{\pm} & =\frac{1}{\beta}\ln\left(n_{\pm}^{{\rm b}}\lambda^{3}\right)\\
\mu_{P} & =\frac{1}{\beta}\ln\left[p_{{\rm b}}(z_{+},z_{-})\lambda_P^{3}\right]+g_{0}\left(z_{+},z_{-}\right),
\end{split}
\end{equation}
where $g_{0}\left(z_{+},z_{-}\right)$ is the bulk value of $g\left(z_{+},z_{-}\right)$, Eq.~(\ref{eq4}), evaluated in the bulk, $\psi=0$. From the relation between $p_{\rm b}(z_{+},z_{-})$ and $\mu_P$, we obtain
\begin{equation} \label{eq7}
p_{\rm b}(z_{+},z_{-})=\mathcal{A}{\rm e}^{-\beta g_0\left(z_{+},z_{-}\right)},
\end{equation}
with the coefficient $\mathcal{A}$ determined by Eq.~(\ref{eq3}) in the bulk, $\sum_{z_{+},z_{-}}p_{\rm b}(z_{+},z_{-})=P_{\rm b}$. Thus, $g\left(z_{+},z_{-}\right)$ assumes its bulk value ($\psi=0$), $g_0 (z_+,z_-)$, that can be written in
the form
\begin{equation} \label{eq8}
\begin{split}\beta g_{0}\left(z_{+},z_{-}\right)= & -\ln\left[{N_{+} \choose z_{+}}{N_{-} \choose z_{-}}\right]+\beta F_{\rm CR}\left(z_{+},z_{-}\right)\\
 & -z_{+}\ln\left(n_{+}^{{\rm b}}\lambda^{3}\right)-z_{-}\ln\left(n_{-}^{{\rm b}}\lambda^{3}\right).
\end{split}
\end{equation}
We note that Eq.~(\ref{eq8}) specifically pertains to the ion {\it association} charging mechanism, {\it i.e.}, {\it adsorption} of the monovalent ions onto the macro-ion. For the 
opposite process of ion {\it dissociation} charging mechanism, Eq.~(\ref{eq8}) would have to be somewhat modified, as is shown in Appendix~\ref{appendixA} for a simple protonation/deprotonation mechanism. For simplicity sake, in the remaining of the paper, we only consider ionization by {\it association} of monovalent ions onto the macro-ion, as in Eq.~(\ref{eq8}).

Finally, thermodynamic equilibrium requires that $\delta F/ \delta \psi=0$. This leads to a generalized Poisson-Boltzmann equation
\begin{equation} \label{eq9}
\begin{split}
&-\varepsilon_
{0}\varepsilon \nabla^2\psi\left({\bf r}\right) = en_+^{\rm b}{\rm e}^{-\beta e\psi\left({\bf r}\right)}-en_-^{\rm b}{\rm e}^{\beta e\psi\left({\bf r}\right)}\\
 & +\sum_{z_{+},z_{-}} e \left(z_{+}-z_{-}\right) p_{\rm b}(z_{+},z_{-}){\rm e}^{-\beta\left(z_{+}-z_{-}\right)e\psi\left({\bf r}\right)}.
\end{split}
\end{equation}
We note that in the bulk, the MF approximation is characterized by a constant electrostatic potential, which was taken here as zero. Therefore, when using the MF theory, the only effect of electrostatics in the bulk is seen in the electro-neutrality condition,
\begin{equation} \label{eq10}
\begin{split}
n_+^{\rm b}+\sum_{z_{+},z_{-}}z_{+} p_{\rm b}(z_{+},z_{-})=n_-^{\rm b}+\sum_{z_{+},z_{-}}z_{-} p_{\rm b}(z_{+},z_{-}).
\end{split}
\end{equation}
%
%%%%%%%%%
\subsection{The CR phenomenological free-energy} \label{model_a}
%%%%%%%%%%
While the system is formally characterized by Eqs.~(\ref{eq5})-(\ref{eq10}), its behavior depends on the phenomenological free-energy $F_{\rm CR}(z_+,z_-)$ that incorporates the details of the CR mechanism. Different {\it CR models} can be implemented by choosing different forms of $F_{\rm CR}$. In the past, we considered only the case when $F_{\rm CR}$ is linear in $z_{\pm}$~\cite{Avni2018, Markovich2018}, resulting in the simplest CR mechanism. This is now extended by taking into account second-order terms in $F_{\rm CR}(z_+,z_-)$ as well. For simplicity, we restrict ourselves to symmetric macro-ions, having ${N \equiv N_+=N_-}$ and a symmetric ${F_{\rm CR}(z_+,z_-)=F_{\rm CR}(z_-,z_+)}$. The more general formulation of the non-symmetric model is presented in Appendix~\ref{appendixB}.

Returning to the symmetric case, $F_{\rm CR}$ has the form
\begin{equation} \label{eq11}
\beta F_{{\rm CR}}=\alpha\left(z_{+}+z_{-}\right)+\frac{\chi_{1}}{2N}\left(z_{+}^{2}+z_{-}^{2}\right)+\frac{\chi_{2}}{N}z_{+}z_{-},
\end{equation}
with dimensionless parameters: $\alpha$, $\chi_1$ and $\chi_2$. The linear term (considered already in Refs.~\cite{Markovich2018, Avni2018}) accounts for the independent free-energy gain from each adsorption, while the quadratic terms represent the change in the free-energy due to short-range interactions between adsorption sites. This is of relevance to CR macromolecules containing dissociable groups where the cooperativity between the surface sites leads to ion adsorption onto different sites that are not mutually independent~\cite{Harries2006, Diamant1996}.

While our model treats the macro-ions as point-like, the interaction terms take into account the fact that sites that are far from each other on the macro-ion surface do not interact. This is accounted by the $1/N$ factor in the interaction terms, which is appropriate for nearest neighbor type of interactions, with $\chi_{1}$ and $\chi_{2}$ being independent of $N$.

If the short-range interaction between equally charged sites is repulsive, and attractive between oppositely charged sites, we have $\chi_{1}>0>\chi_{2}$, while $\chi_{2}>0>\chi_{1}$ is given by the opposite case.
We note that the sign and strength of $\chi_{1,2}$ does not necessarily relate to simple attraction/repulsion between association sites, but can also indicate changes in the conformational free-energies or the stability of the charge states. In either case, we treat $\chi_{1,2}$ as phenomenological parameters.

The symmetry properties of the CR model, together with the condition for overall neutrality, implies ${n_{+}^{\rm b}=n_{-}^{\rm b}\equiv n_{\rm b}}$. However, in an inhomogeneous system, {\it e.g.}, in the presence of an external electric field, the symmetry between the positive and negative local concentrations will be broken.
%%%%%%%%
\subsection{Bulk CR behavior: The large $N$ limit} \label{model_b}
%%%%%%%%%%%

We now consider the limit of large number of adsorption sites, $N \gg 1$. It is then convenient to refer to the charge state of each macro-ion not by $z_+$ and $z_-$, but by the fractions of positively and negatively charged sites, defined as $\phi_{+}\equiv z_{+}/N$ and $\phi_{-}\equiv z_{-}/N$ with $0\leq\phi_{\pm}\leq1$.

Defining $\tilde{g}_0\left(\phi_{+},\phi_{-}\right)=g_0\left(z_{+},z_{-}\right)/N$ and using the Stirling's formula, we obtain
\begin{equation} \label{eq12}
\begin{split}\beta\tilde{g}_{0}\left(\phi_{+},\phi_{-}\right)= & \phi_{+}\ln\phi_{+}+\left(1-\phi_{+}\right)\ln\left(1-\phi_{+}\right)\\
 & +\phi_{-}\ln\phi_{-}+\left(1-\phi_{-}\right)\ln\left(1-\phi_{-}\right)\\
 & -\left(\phi_{+}+\phi_{-}\right)\left[\ln\left(n_{{\rm b}}\lambda^{3}\right)-\alpha\right]\\
 & +\frac{1}{2}\chi_{1}\left(\phi_{+}^{2}+\phi_{-}^{2}\right)+\chi_{2}\phi_{+}\phi_{-}.
\end{split}
\end{equation}
where $\tilde{g}_0$ is defined up to a constant independent of $\phi_{\pm}$. Substituting Eq.~(\ref{eq12}) in Eq.~(\ref{eq7}) we obtain
\begin{equation} \label{eq13}
p_{\rm b}(\phi_{+},\phi_{-})=\mathcal{A}{\rm e}^{-\beta\tilde{g}_0\left(\phi_{+},\phi_{-}\right)N}.
\end{equation}

For fixed interaction parameters $\alpha, \chi_1, \chi_2$ and a fixed bulk ion concentration $n_{\rm b}$, the distribution function $p_{\rm b}(\phi_{+},\phi_{-})$ is characterized by a sharp peak at the minimum of $\tilde{g}_0\left(\phi_{+},\phi_{-}\right)$. If $\tilde{g}_0$ has a single global minimum, all the macro-ions (except for a negligible fraction), will be characterized by the fractions $\phi_{\pm}$ calculated at the global minimum. The fractions $\phi_{\pm}$ that extremize $\tilde{g}_0$ are obtained from the two equations
\begin{equation} \label{eq14}
\begin{split}\frac{\phi_{+}}{1-\phi_{+}} & =n_{{\rm b}}\lambda^{3}{\rm e}^{-\alpha-\chi_{1}\phi_{+}-\chi_{2}\phi_{-}}\\
\frac{\phi_{-}}{1-\phi_{-}} & =n_{{\rm b}}\lambda^{3}{\rm e}^{-\alpha-\chi_{1}\phi_{-}-\chi_{2}\phi_{+}}.
\end{split}
\end{equation}
We note that if ${\chi_2=0}$, {\it i.e.}, there is no interaction between the different types of sites, and the equations above reduce to two Langmuir-Frumkin-Davies~\cite{Markovich2018,Harries2006,Tomer2020} adsorption isotherms,
\begin{equation} \label{eq15}
\begin{split}\phi_{\pm} & =\frac{1}{1+\left(n_{{\rm b}}\lambda^{3}\right)^{-1}{\rm e}^{\alpha+\chi_{1}\phi_{\pm}}}.
\end{split}
\end{equation}
However, $\tilde g_0$ may have more than one minimum, depending on the values of system parameters, and under certain conditions, $\tilde g_0$ can have more than one {\it global} minimum, leading to a coexistence of different types of macro-ion charge states.

We define $\langle...\rangle_{{\rm b}}$ to be the average over all charge configurations of macro-ions in the bulk,
\begin{equation} \label{eq16}
\langle...\rangle_{{\rm b}}=\frac{1}{P_{{\rm b}}}\sum_{z_{+},z_{-}}\left(...\right)p_{{\rm b}}\left(z_{+},z_{-}\right).
\end{equation}
Due to the symmetry of our model, $\langle\phi_{+}\rangle_{\rm b}=\langle\phi_{-}\rangle_{\rm b}$. However, this does not necessarily mean that the macro-ions are overall neutral. As an example, a solution where half of the macro-ions have a net charge of $Q$, and the other half $-Q$, satisfies the above symmetry condition. In order to better understand the macro-ion charge states, it is useful to define two order parameters,
\begin{equation} \label{eq17}
\begin{split}
\Phi & \equiv\frac{1}{2}\langle \phi_{+}+\phi_{-}\rangle_{\rm b} \\
Z & \equiv\sqrt{\langle \left(\phi_{+}-\phi_{-}\right)^{2}\rangle_{\rm b}}\,\,\,,
\end{split}
\end{equation}
where $0 \leq \Phi \leq 1$ is the total fraction of charged sites, and $Z$ is the standard deviation of the macro-ion charge divided by $N$. Both $\Phi$ and $Z$ are limited to the $[0,1]$ range, where $\Phi=0$ (or $1$) corresponds to macro-ions with completely empty (or full) sites, and $Z=0$ (or $1$) corresponds to neutral (or maximally charged) macro-ions. In the $N\gg 1$ limit, the $(\Phi_,Z)$ state corresponds to the global minimum/minima of $\tilde{g}_0$, which is determined by $\chi_{1,2}$ and by the combination $\ln(n_{\rm b}\lambda^3)-\alpha$ (see Eq.~(\ref{eq12})). Defining the rescaled chemical potential,
\begin{equation} \label{eq18}
\mu' \equiv \ln(n_{\rm b}\lambda^3) -\alpha,
\end{equation}
we conclude that the bulk system can be described by a 3D phase diagram, showing the $(\Phi_,Z)$ state as a function of $\chi_1$, $\chi_2$ and $\mu'$. {\footnote{Throughout the paper $\alpha$ is taken as a phenomenological parameter. In a microscopic model that describes the site potential, the entropic part of $F_{\rm CR}(z_+,z_-)$, and therefore $\alpha$ as well, depend on $\lambda$ in a way that cancels the dependence of $\mu'$ on $\lambda$. Thus, the choice of $\lambda$ as the length scale is arbitrary here.}

%%%%%%%
\section{Results and Discussion} \label{Results}
%%%%%%%%%%%%%
%%%%%%%
\subsection{The phase space} \label{results_a}
%%%%%%%%%%%%%

 %%%%%%%%%
%Fig2
%%%%%%%%%%%%%%%%%%%%%%%%%%%%%%%%%%%%%%%%%%%
\begin{figure*}
\includegraphics[width = 2\columnwidth,draft=false]{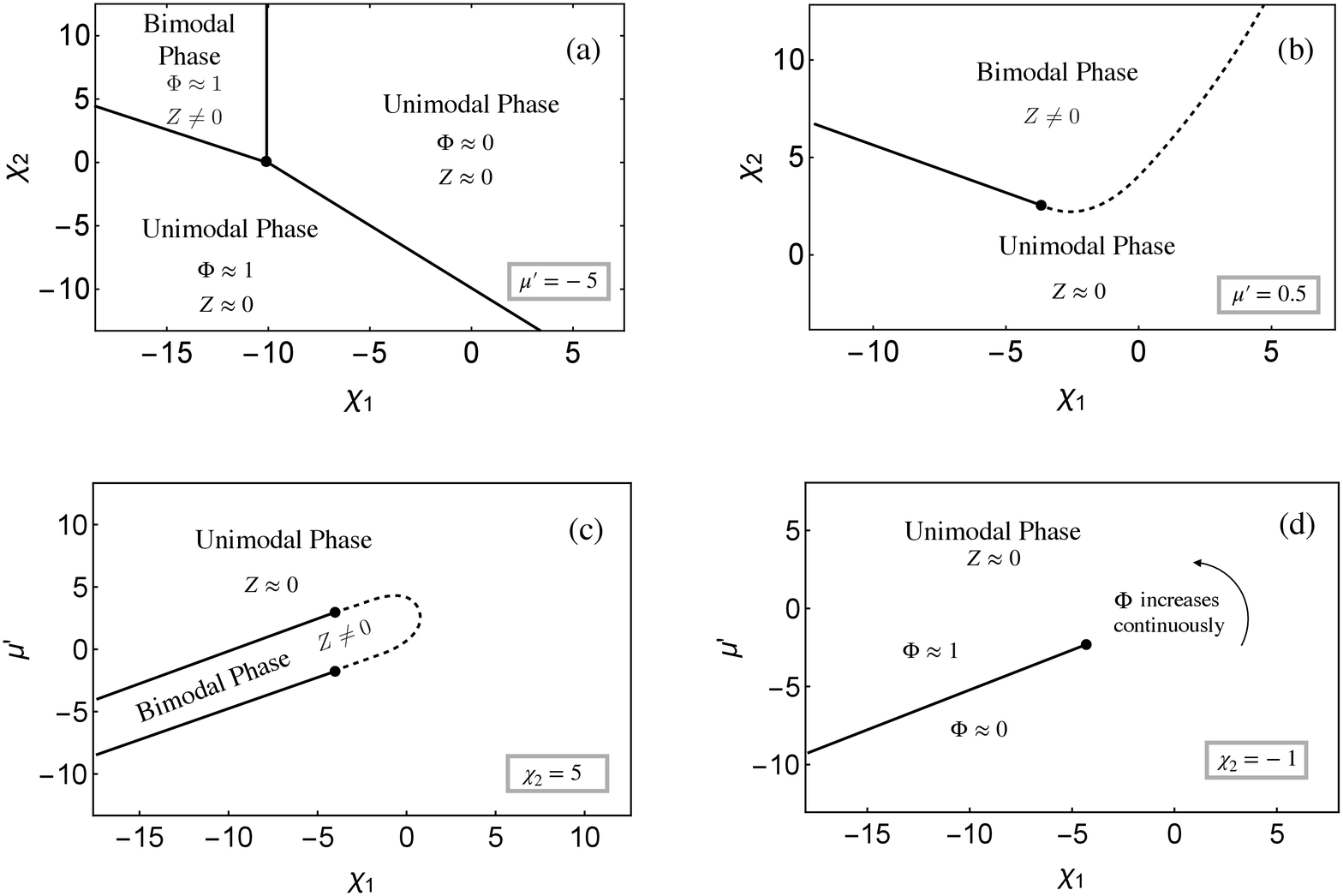}
\caption{\textsf{(Color online) Cuts through the $(\chi_1,\chi_2,\mu')$ 3D phase-diagram of macro-ions with (a) ${\mu'=-5}$, (b) $\mu'= 0.5$, (c) ${\chi_2 = 5}$ and (d) ${\chi_2 = -1}$. The interaction between similar and different macro-ion sites are given by $\chi_1$ and $\chi_2$, respectively, and $\mu'$ is related to the bulk ion concentration and to the non-interacting part of the free-energy gain by adsorption, as in Eq.~(\ref{eq18}). The different phases are separated by the $(\Phi,Z)$ values (as defined in Eq.~(\ref{eq17})), and are set by the global minimum of $\tilde{g}_0$. A unimodal phase corresponds to a single global minimum, while a bimodal phase to two global minima. The phases are separated by a solid line if $\Phi$ or $Z$ changes discontinuously (due to an emergence of a new global minimum/minima), and by a dashed line if both $\Phi$ and $Z$ change continuously (due to a bifurcation or merger of the global minimum/minima).}}
\label{Fig2}
\end{figure*}
%%%%%%%%%%%%%%%%%%%%%%%%%%%%%%%%%%%%%%%%%%%

We construct a phase diagram (Fig.~\ref{Fig2}) that distinguishes various $\{\Phi,Z\}$ states, depending on the values of $\chi_1$, $\chi_2$ and $\mu'$. Figure~\ref{Fig2} displays cuts through the full 3D phase diagram, with $\mu'={\it const.}$ in (a) and (b), and $\chi_2={\it const.}$ in (c) and (d). Full lines represent first-order phase transitions, where a new global minimum/minima of $\tilde{g}_0$ emerges, and consequently the derivative of the free energy with respect to $\chi_1$ and $\chi_2$ in (a) and (b) and $\chi_1$ and $\mu'$ in (c) and (d) is discontinuous. Dashed lines describe second-order phase transition, where there is a bifurcation or merge of the global minimum/minima, leading to continuous first derivatives of the free energy, but discontinuous second derivatives. We note that the objective in the following analysis is to show representative cases of phase transitions through 2D cuts in the phase diagram, rather than presenting the full phase diagram.

For $\mu'=-5$ (Fig.~\ref{Fig2} (a)), we obtain three distinct phases: (i) a {\it unimodal phase} where $\tilde{g}_0$ has a single global minimum, with $(\Phi\approx 0,Z\approx 0)$; (ii) a {\it unimodal phase} with $(\Phi\approx 1,Z\approx 0)$; and (iii) a {\it bimodal phase} where $\tilde{g}_0$ has two  global minima with $(\Phi\approx 1,Z>0)$. In the first phase, the macro-ions are overall neutral. Moreover, most of their sites are neutral. This phase exists for large $\chi_1$ and $\chi_2$. The second phase describes overall neutral macro-ions, having charged sites that roughly cancel each other. Such a situation occurs for $\chi_1$ and $\chi_2$ at the left bottom corner of Fig.~\ref{Fig2}(a) (either negative or positive and small). The third phase corresponds to a system with two types of macro-ions: highly positively charged or highly negatively charged, occurring for small $\chi_1$ and large $\chi_2$, {\it i.e.}, attraction between same charge site, and repulsion between opposite charge sites. As the system crosses the first-order phase-transition line (drawn as solid black line), the global minimum changes discontinuously between two (or more) local minima, causing either $\Phi$ or $Z$ to experience a ``jump". The three phases meet at $\chi_1\approx-10$ and $\chi_2\approx0$. At this point, there is in fact a coexistence of four macro-ion charge states, because the binodal phase contains two charge states. We note that this is not a regular ``triple point", as it extends in the 3D phase diagram to a line rather than a point.

For $\mu'=0.5$ (Fig.~\ref{Fig2}(b)), the phase diagram is significantly different from that of~Fig.~\ref{Fig2}(a) with $\mu'=-5$. Here there are only two phases: (i) a unimodal phase with $Z\approx0$, describing a system with one dominant type of macro-ions that are overall neutral; and (ii) a bimodal phase with $Z\neq0$, corresponding to a system with two types of macro-ions: positively charged and negatively charged. Unlike the case shown in Fig.~{\ref{Fig2}(a)}, the transition here from unimodal to bimodal phase can occur in two ways. For large $\chi_1$, it occurs through the bifurcation of a single global minimum into two, causing $Z$ to increase {\it continuously} from zero (dashed line in Fig.~\ref{Fig2}(b)). In the second case of negative and large $\chi_1$, it occurs via an emergence of two new local minima, which at some point surpass the previous global minimum, causing $Z$ to ``jump" from zero to a finite value (solid line in Fig.~\ref{Fig2}(b)).

In Fig.~\ref{Fig2}(c), the phase diagram in the $\left(\mu', \chi_1\right)$ plane is shown for the case of repulsion between oppositely charged sites, $\chi_2>0$, while in~\ref{Fig2}(d), it is shown for the attraction case, $\chi_2<0$. We note that as $\mu'$ is a function of the ionic concentrations, changing $\mu'$ is similar to changing the pH (see Appendix~\ref{appendixA} for the explicit relation between our model parameters and the pH and pK$_a$ in a protonation/deprotonation mechanism). In both figure parts~\ref{Fig2}(c) and~\ref{Fig2}(d), the phase-transition line terminates at an end-point in the $(\mu',\chi_1)$ plane.
Hence, changing $\mu'$ for negative and large $\chi_1$ values leads to a phase transition (``jump" in either $\Phi$ or $Z$), whereas at sufficiently large $\chi_1$, a change in $\mu'$ leads to a continuous change (no transition) in $\Phi$, together with $Z=0$. We further note on the difference between (c) and (d). In (c) the transition between two unimodal phases is pre-empted by a bimodal phase, whereas in (d), the system ``jumps" directly from one unimodal phase to the second one.

For a given macro-ionic solution, the interaction parameters are fixed and the only parameter that varies in the phase space described by Fig.~\ref{Fig2} is, $n_{\rm b}$, the bulk concentration of salt, which is induced by a change in $\mu'$. From the analysis presented above (Fig.~\ref{Fig2}), we conclude that changing the salt concentration can vary the macro-ion charge state, and particularly, it can shift the macro-ion state from a unimodal charge distribution to a bimodal one. A bimodal phase that persists for a large range of $n_{\rm b}$ (as in Fig.~\ref{Fig2}(a)-(c)) is unique for the symmetric model. The multi-dimensional phase-space in a non-symmetric model might have multi-modal phases as well, but of a more complicated form (see Appendix~\ref{appendixB}).

 %%%%%%%%%%%%
%Fig3
%%%%%%%%%%%%%%%%%%%%%%%%%%%%%%%%%%%%%%%%%%%
\begin{figure*}
\includegraphics[width = 2\columnwidth,draft=false]{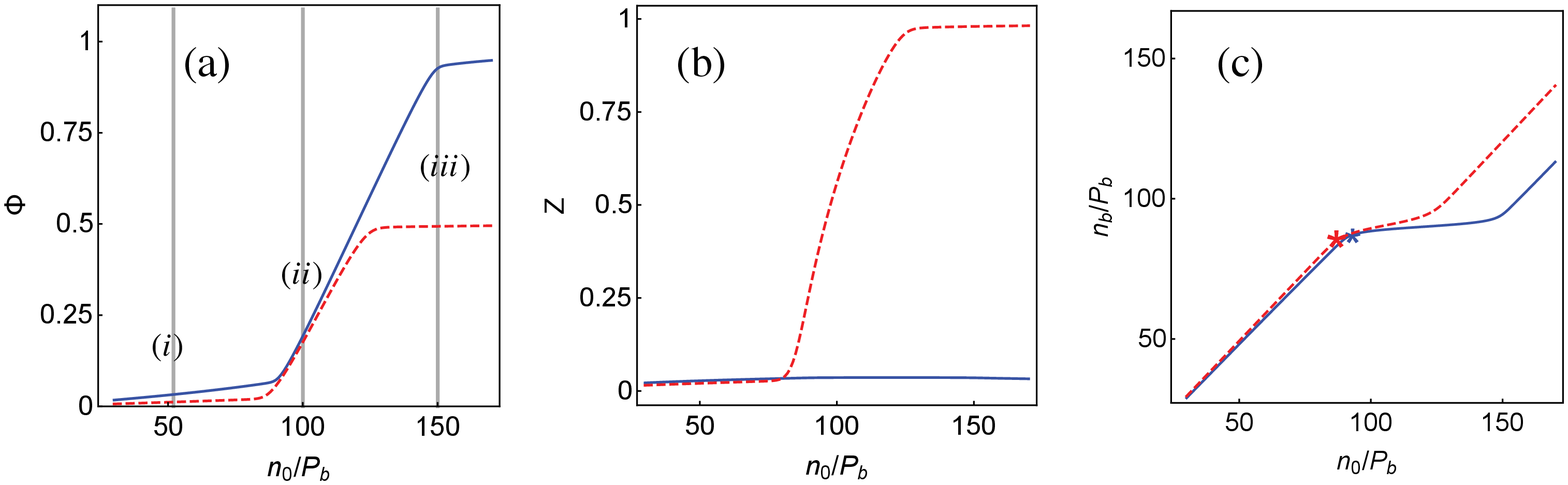} %1
\caption{\textsf{(Color online) The parameters $\Phi$, $Z$ and $n_{\rm b}/P_{\rm b}$ are shown as functions of the total added salt concentration $n_0$, normalized by a fixed macro-ion concentration, $P_{\rm b}$, in (a), (b) and (c), respectively. The interaction parameters are $\chi_1=2$, $\chi_2=-8$ and $\mu''=-7.5$ for the solid blue line, and $\chi_1=-8$, $\chi_2=2$ and $\mu''=-8.5$ for the dashed red line. For both plotted cases, the number of positive and negative sites is taken as $N=60$. The gray vertical lines in (a) correspond to the concentrations for which the macro-ion state distribution is presented  in Fig.~\ref{Fig4}. The onset of the plateau in (c), which defines the $n_0^{\rm CIC}$, is highlighted by a star marker.}}
\label{Fig3}
\end{figure*}
%%%%%%%%%%
%%%%%%%%%%%%%%%%%%%%%%%%%%%%%%%%%
 %%%%%%%%%%%%
%Fig4
%%%%%%%%%%%%%%%%%%%%%%%%%%%%%%%%%%%%%%%%%%%
\begin{figure*}
\includegraphics[width = 1.5\columnwidth,draft=false]{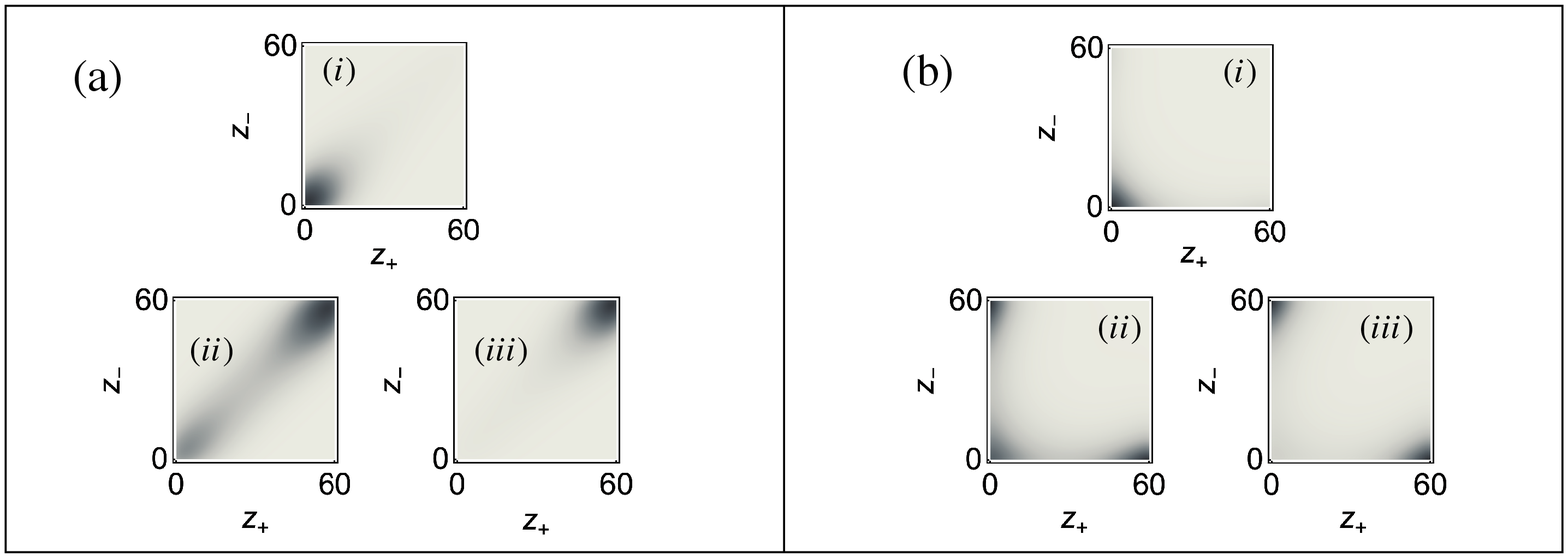} 
\caption{\textsf{The distribution of macro-ions  $p_{\rm b}(z_+,z_-)$ at each $(z_{+},z_{-})$ state. In (a) the macro-ion parameters are $\chi_1=2$, $\chi_2=-8$, $\mu''=-7.5$ and $N=60$, corresponding to the solid blue line in Fig.~\ref{Fig3}. In (b) the macro-ion parameters are $\chi_1=-8$, $\chi_2=2$, $\mu''=-8.5$ and $N=60$, corresponding to the red dashed line in Fig.~\ref{Fig3}. The distribution is shown for three values of the total salt concentrations, indicated in Fig.~\ref{Fig3}(a) by vertical gray lines, (i) $n_0=50 P_{\rm b}$, (ii) $n_0=100 P_{\rm b}$ and (iii) $n_0=150 P_{\rm b}$. The gray color code is associated with the magnitude of $p_{\rm b}(z_+,z_-)$}.}
\label{Fig4}
\end{figure*}
%%%%%%%%%%
%%%%%%%%%%
%%%%%%%%%%%
%Fig5
\begin{figure}
\includegraphics[width = 1\columnwidth,draft=false]{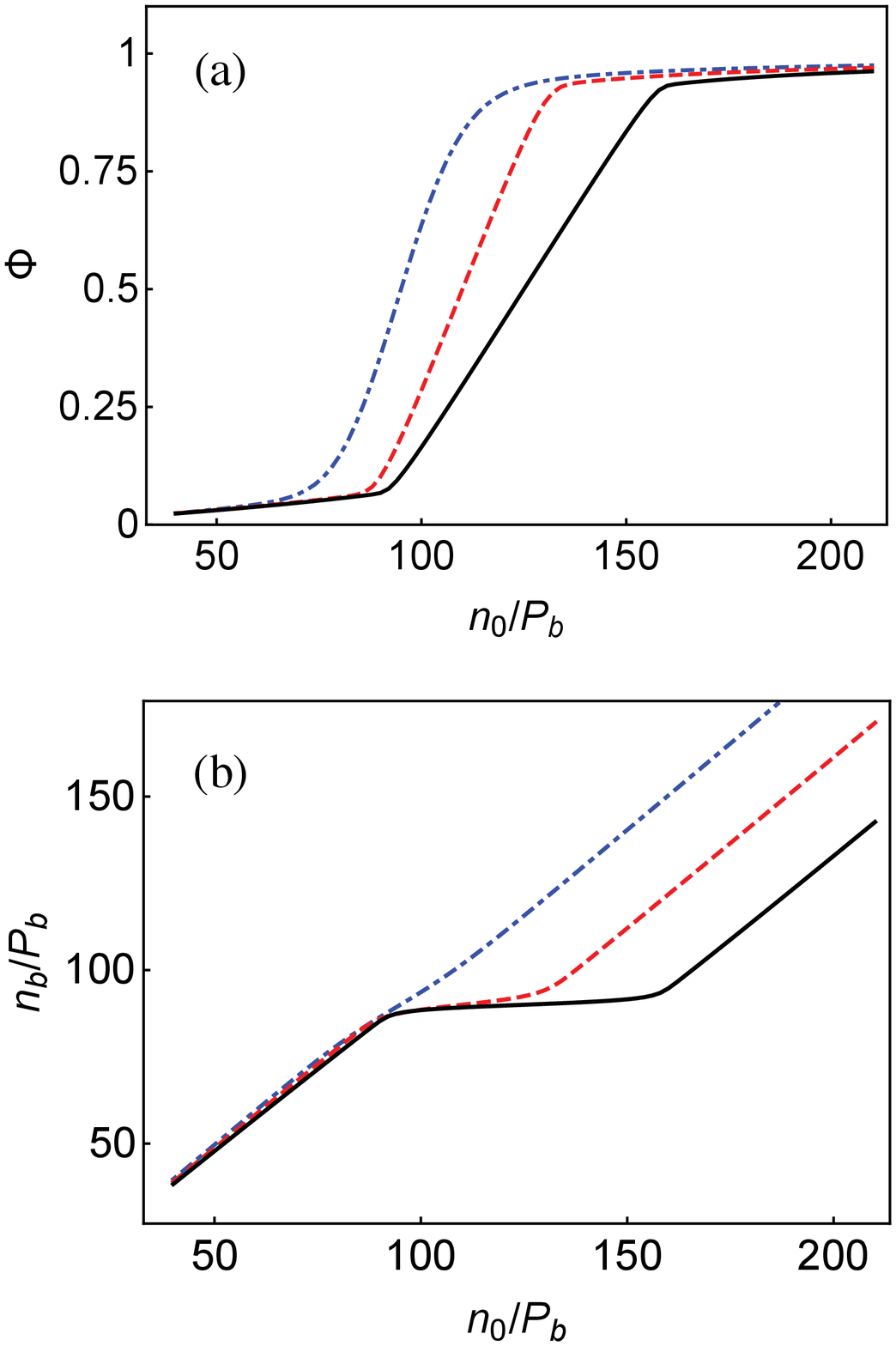} 
\caption{\textsf{(Color online) (a) The order parameter $\Phi$, and (b) $n_{\rm b}/P_{\rm b}$ as function of the total added salt concentration $n_0$, normalized by fixed macro-ion concentration $P_{\rm b}$, for $N=10$ (dotted-dashed blue line), $N=40$ (dashed red line) and $N=70$ (solid black line). The interaction parameters are $\chi_1=2$, $\chi_2=-8$ and $\mu''=-7.5$. As $N$ increases, the linear increase of $\Phi$ and the plateau of $n_{\rm b}$, when $n_0>n_0^{\rm CIC}$, become sharper.}}
\label{Fig5}
\end{figure}
%%%%%%%%%%%%%%%%%%%%%%%%%%%%%%%%%%%%%%%%%%%%

%%%%%%%
\subsection{The bulk behavior close to the critical salt concentration} \label{results_b}
%%%%%%%%%%%%%
In the previous section, we investigated the phase space in the $N\gg1$ limit, and did not take into account the explicit relationship between the macro-ion concentration and the bulk ion concentration, $n_{\rm b}$, given by
\begin{equation} \label{eq19}
n_{0}=n_{\rm b}+\sum_{z_{+},z_{-}}z_{\pm} p_{\rm b}(z_{+},z_{-}),
\end{equation}
where $n_0$ is the total concentration of positive/negative ions (which are equal due to the electro-neutrality condition (Eq.~(\ref{eq10})). Note that the parameter $n_0$ should be distinguished from $n_{\rm b}$, because the latter $n_{\rm b}$ parameter is the concentration of free ions in the solution, which are not adsorbed by the macro-ions. As the macro-ions are added to the solution in a neutral state, and adsorb ions from the solution, the former $n_0$ (rather than $n_{\rm b}$) is our control parameter, as it corresponds to the total added salt concentration.

We recall that the concentration of macro-ions at each charge state, $p_{\rm b}(z_{+},z_{-})$, is determined by $n_{\rm b}$ through Eqs.~(\ref{eq7}) and~(\ref{eq8}). Thus, via $p_{\rm b}$, Eq.~(\ref{eq19}) represents a complicated relationship  between $n_0$ and $n_{\rm b}$. However, as the relation is monotonic it can be inverted, allowing the calculation of the free ion concentration, macro-ion concentration, and the order parameters $\Phi$ and $Z$, as a function of $n_0$. Assuming that the concentration of macro-ions, $P_{\rm b}$, is fixed, while $n_{\rm b}$ changes as function of $n_0$, it is more convenient to define a new rescaled chemical potential to be
\begin{equation} \label{eq20}
\mu'' \equiv \ln(P_{\rm b}\lambda^3)-\alpha=\mu'-\ln(n_{\rm b}/P_{\rm b}).
\end{equation}
Note that $\mu''$ is held fixed while changing $n_0$. We recall that based on the analysis of the $N\gg1$ limit in the previous section, at some $n_{\rm b}$, and for certain interaction parameters, the system crosses a first-order phase transition line where the global minimum/minima of $\tilde{g}_0$ shifts from one or two sets of $(\phi_+,\phi_-)$ values to another one/two sets. As $n_0$, rather than $n_{\rm b}$, is our controlled parameter, the shift between global minima of $\tilde{g}_0$ is expected to start at some critical $n_0$.

Figure~\ref{Fig3} (a), (b) and (c) shows the behavior of $\Phi$, $Z$ and $n_{\rm b}$, respectively, as function of $n_0$, for different macro-ion parameters $\chi_1$, $\chi_2$, $\mu''$, and for $N=60$. The solid (blue) line depicts the case where $\chi_1>0$ and $\chi_2<0$, {\it i.e.}, {\it repulsion} between sites having the same charge and {\it attraction} between oppositely charged sites. The dashed (red) line shows the opposite case, of attraction between sites having the same charge, $\chi_1<0$, and repulsion between oppositely charge sites, $\chi_2>0$. The distribution of macro-ion charge states, $p_{\rm b}(z_+,z_-)$ at three different $n_0$ concentrations, depicted by gray vertical lines in Fig.~\ref{Fig3}(a), are shown in Fig.~\ref{Fig4}, and further discussed below. In the following, we denote the critical ionic concentration, $n_0^{\rm CIC}$, as the concentration that corresponds to the onset of the plateau in the $n_{\rm b}$ plot (see star markers in Fig.~\ref{Fig3}(c)).
%\newline
\subsubsection{The case of $\chi_1>0$ and $\chi_2<0$}
For low salt concentrations, the macro-ions adsorb a small amount of positive and negative ions, such that both $\Phi\approx0$ and $Z\approx0$ . Around the critical ionic concentration $n_0^{\rm CIC}\simeq 95 P_{\rm b}$ (see solid blue line in Fig.~\ref{Fig3}(a)), similar to the critical micelle concentration (CMC)~\cite{Israelachvili1992}, it becomes favorable for the macro-ions to adsorb many ions and to reach almost full occupancy of their sites. By adding more salt, beyond the critical concentration, $n_0>n_0^{\rm CIC}$, the additional ions get adsorbed by some of the macro-ions, resulting in two-phase coexistence, where macro-ions with low occupation fractions $\phi_{\pm}\approx 0$, coexist with macro-ions with high values, $\phi_{\pm}\lesssim1$ (see Fig.~\ref{Fig4}(a)). We note that the limiting cases of $\phi_{\pm}=1$ or $0$ are never reached due to the gain in entropy of mixing.

Upon increasing the salt concentration $n_0$, the concentration of highly adsorbing macro-ions increases at the expense of the concentration of weakly adsorbing macro-ions, until
almost all of the macro-ions are in the high adsorption state, and any additional ions remain free in the solution. This process results in two distinct behaviors: (i) $\Phi$ increases linearly from a small to a large value, Fig.~\ref{Fig3}(a), and (ii) $n_{\rm b}$ goes through a plateau until it rises again (Fig.~\ref{Fig3}(c)). As for $Z$, it slightly increases in the coexistence region and then decreases again. However, it does not go through a noticeable change, Fig.~\ref{Fig3}(b), meaning that the macro-ions continue to be overall neutral.
\subsubsection{The case of $\chi_1<0$ and $\chi_2>0$}
%\newline
In a similar way to the previous case, there is a critical ionic concentration $n_0^{\rm CIC}\simeq 85 P_{\rm b}$, above which another population of macro-ions develops with high occupation fraction $\Phi$ (dashed blue line in Fig.~\ref{Fig3}(a)), leading to a plateau in $n_{\rm b}$ (Fig.~\ref{Fig3}(c)). However, this population is characterized by highly charged macro-ions, having either saturated positive adsorption, or negative ones, but not both (Fig.~\ref{Fig4}(b)), corresponding to the bimodal phase introduced in Sec.~\ref{results_a} (see Fig.~\ref{Fig2}). Therefore, upon increasing $n_0$ beyond the critical concentration, $Z$ increases until it reaches $Z \approx 1$. Then, the bimodal population of highly charged macro-ions takes over (Fig.~\ref{Fig3}(b)) and $\Phi\simeq0.5$ (Fig.~\ref{Fig3}(a)).
\newline

We further comment on both cases discussed above. The results presented in Figs.~\ref{Fig3} and~\ref{Fig4} are for a large number of sites, $N=60$. For smaller values of $N$, a similar transition occurs for the macro-ion charge states, but its effect on the order parameter is different. Figure~\ref{Fig5} shows $\Phi$ and $n_{\rm b}$ as function of $n_0$ for $N=10$, $40$ and $70$. For simplicity, it shows a case with $\chi_1>0$ and $\chi_2<0$ case (the opposite case of $\chi_1<0$ and $\chi_2>0$ is qualitatively the same). It is evident that as $N$ decreases, the transition of $\Phi$ becomes less sharp and the plateau of $n_{\rm b}$ loses its flatness and becomes narrower, until it is not seen at all for $N=10$.

As seen above, the nucleation of additional macro-ion charge states resulted from surpassing a critical salt concentration, $n_0>n_0^{\rm CIC}$, while keeping the concentration of macro-ions fixed. However, the criticality is determined by the ratio between the total salt concentration and the macro-ion concentration, $n_0/P_{\rm b}$. Therefore, we can alternatively vary $P_{\rm b}$, while fixing $n_{\rm 0}$, and obtain, in complete analogy, a critical macro-ion concentration.

We conclude by comparing the above results to those presented in Sec.~\ref{results_a}. The macro-ion charge state was shown in Fig.~\ref{Fig2} to experience, in some cases, a ``jump" when changing $n_{\rm b}$, suggesting a first-order phase transition. However, we have seen as well that this is an artifact of changing $n_{\rm b}$, rather than $n_{0}$. When the latter $n_0$ is changed, the transition is never abrupt but changes continuously as shown in Figs.~\ref{Fig3} and~\ref{Fig4}. When $n_0>n_0^{\rm CIC}$, $n_{\rm b}$ experiences a plateau, which is why the transition regions appear as coexistence lines in~Fig.~\ref{Fig2}. Moreover, in the transition region a bimodal, and even a trimodal phase can develop (see Fig.~\ref{Fig4}(b) where macro-ions with high positive, high negative, and neutral charge coexist). Note that the multimodal phases in the transition between two free-energy minima are not unique to the symmetric model but rather are expected to occur in a general non-symmetric model.
%%%%%%%%%%%%%%%%%
\subsection{Externally applied electric fields} \label{results_c}
%%%%%%%%%%%%%%%%%%%
Until now, we analyzed the bulk properties of a homogeneous macro-ion solution. In order to investigate interfacial properties, we consider an inhomogeneous system in presence of a planar surface held at fixed potential $V$, at $x=0$. The electrostatic potential, $\psi(x)$, has a spatial variation close to the interface, and the corresponding PB equation, Eq.~(\ref{eq9}), should be solved with the boundary conditions $\psi(0)=V$ and $\psi(\infty)=\psi'(\infty)=0$. At small potentials, {\it i.e.} $e z \psi/k_{\rm B} T\ll 1$ for all possible macro-ion valencies $z$, the above equation can be linearized, and the resulting potential decays exponentially with an effective screening length,
\begin{equation} \label{eq21}
\lambda_{\text{eff}}^{-2} =4 \pi l_{\rm B}
\left[n_{+}^{\rm b}+n_{-}^{\rm b}+\sum_{z_{+},z_{-}} (z_{+}-z_{-})^{2}p_{\rm b}(z_{+},z_{-})\right],
\end{equation}
where $l_{\rm B}$ is the Bjerrum length, $l_{\rm B}=e^2/(4\pi \varepsilon_0 \varepsilon k_{\rm B}T)$.

Within the symmetric model, we obtain
\begin{equation} \label{eq23}
\lambda_{\text{eff}}  =\big[ 4 \pi l_{\rm B}\left(2n_{\rm b}+N^2 Z^2 P_{\rm b}\right)\big]^{-1/2}.
\end{equation}
As $\lambda_{\text{eff}}$ depends on both $Z$ and $n_{\rm b}$, it shows a rather complicated variation, see Fig.~\ref{Fig3}. This is reflected in the dependence of the effective screening length $\lambda_{\text{eff}}$ on $n_0$. Note that in the absence of macro-ions, $P_{\rm b}=0$, $n_{\rm b}=n_0$, and the screening length reduces to the standard Debye screening length, $\lambda_{\rm D}=\left(8\pi l_{\rm B}n_0\right)^{-1/2}$.

In Fig.~\ref{Fig6}, we show the dependence of $\lambda_{\text{eff}}$ on $n_0$, at fixed macro-ion concentration $P_{\rm b}$, for the same interaction parameters as the two cases considered in Fig.~\ref{Fig3}. In the case of repulsion between sites having the same charge ($\chi_1>0$) and attraction between oppositely charged sites ($\chi_2<0$), $\lambda_{\text{eff}}$ goes through a plateau together with $n_{\rm b}$ (see solid blue line in Fig.~\ref{Fig3}(c)), as $Z$ barely changes. In the opposite case, $\chi_1<0$ and $\chi_2>0$, representing attraction between sites having the same charge, and repulsion between oppositely charge sites, the effective screening length $\lambda_{\text{eff}}$ decreases substantially, due to an increase in $Z$ (see dashed red line in Fig.~\ref{Fig3}(b)), and  the system crosses over from having a large screening length, to having a very small one.

%%%%%%%%%%%%
%Fig6
%%%%%%%%%%%%%%%%%%%%%%%%%%%%%%%%%%%%%%%%%%%
\begin{figure}
\includegraphics[width = 1\columnwidth,draft=false]{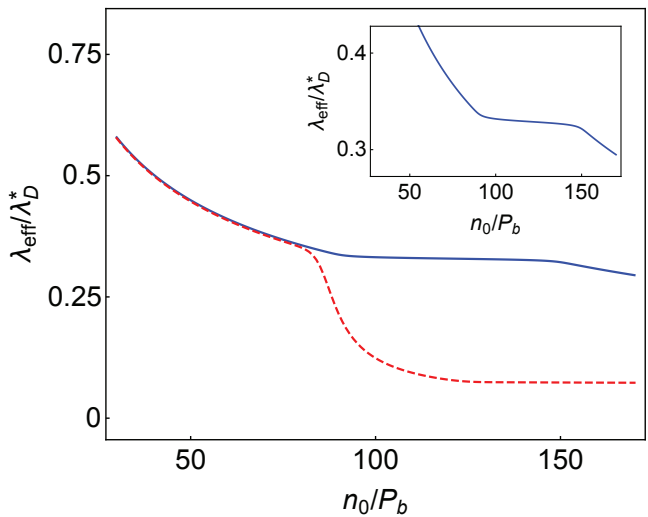} %0.5
\caption{\textsf{(Color online)} The effective screening length, $\lambda_{\text{eff}}$, normalized by $\lambda_{\rm D}^*$, where $\lambda_{\rm D}^*$ is the Debye screening length, $\lambda_{\rm D} = (8\pi l_{\rm B} n_0)^{-1/2}$ evaluated at $n_0/P_{\rm b}=10$, as a function of the total salt concentration $n_0$, normalized by the total macro-ion concentration, $P_{\rm b}$. The interaction parameters are, $\chi_1=2$, $\chi_2=-8$ and $\mu''=-7.5$ (solid blue line) and $\chi_1=-8$, $\chi_2=2$ and $\mu''=-8.5$ (dashed red line), as in Fig.~\ref{Fig3}. In both (a) and (b) $N=60$. In the inset, the solid blue line is shown, at a different $\lambda_{\text{eff}}$ scale, in order to show more clearly the plateau.}
\label{Fig6}
\end{figure}
%%%%%%%%%%%%%%%%%%%%%%%%%%%%%%%%%%%%%%%%%%

For the non-linear PB regime, Eq. (\ref{eq9}) can be solved numerically, and the distribution of ions $n_{\pm}(x)$ and macro-ions $p(z_{\pm};x)$ near the surface can be analyzed. 
In Fig.~\ref{Fig7}, we show the concentration of macro-ions near an electrode held at a negative surface potential, $V<0$, as function of the total charge $Q=e\left(z_+-z_-\right)$, for two sets of macro-ion parameters.
The charge distribution of $p(Q)$, at a given $x$, either has a single peak that shifts towards $Q=0$ as $x$ grows (Fig.~\ref{Fig7} (a)), or several distinct peaks that change their relative hight between one another as function of position (Fig.~\ref{Fig7} (b)). In the latter case (Fig.~\ref{Fig7} (b)), it is shown that for $x<\lambda_{\rm eff}$, a substantial fraction of the macro-ions are charged with $Q/e>10$, while for $x>\lambda_{\rm eff}$, most of the macro-ions are overall neutral.

Comparing this result with the standard double-layer density profile for a simple salt (absence of the macro-ions), the presence of an electrified surface in the macro-ion solution thus leads to a segregation of the different charged species as a function of the distance from the surface, $x$, of order $\lambda_{\rm eff}$.

%%%%%%%%%%%%
%Fig7
%%%%%%%%%%%%%%%%%%%%%%%%%%%%%%%%%%%%%%%%%%%
\begin{figure}
\includegraphics[width =1\columnwidth,draft=false]{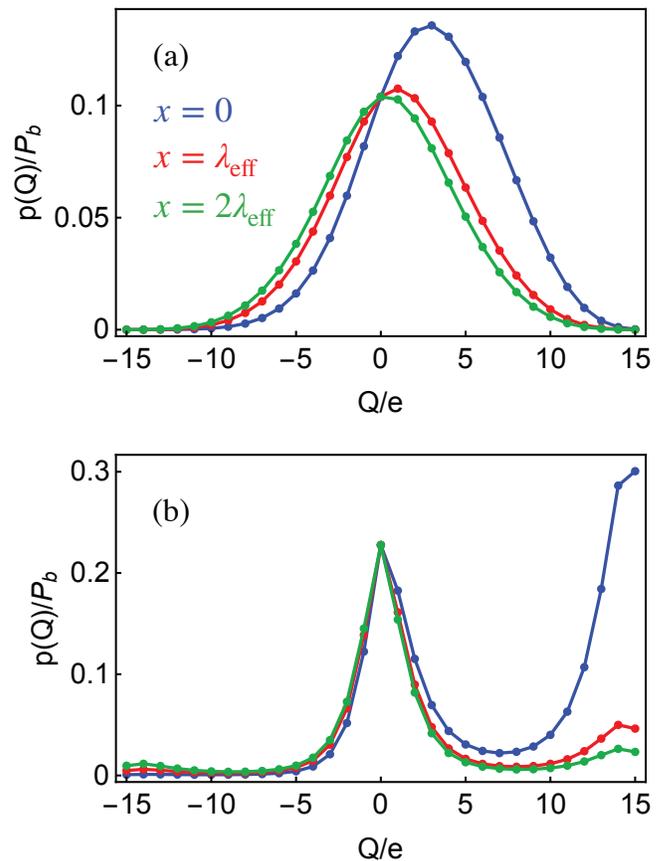} %0.5
\caption{\textsf{(Color online)} The concentration, $P(Q)$, of macro-ions at a fixed distance from an electrode with fixed potential, normalized by the total macro-ion concentration, $P_{\rm b}$, as a function of the total charge $Q/e$.
In (a) the parameters are $\chi_1=-3$, $\chi_2=0$, $\mu''=-8$ and $n_{\rm 0}=500 P_{\rm b}$. In (b) $\chi_1=-6$, $\chi_2=2$, $\mu''=-8$ and $n_{\rm 0}=130 P_{\rm b}$. In both (a) and (b), $V=-0.2k_{\rm B}T/e$ and $N=15$. The blue, red and green lines show the distributions at $x=0,\lambda_{\rm eff}$ and $2\lambda_{\rm eff}$, respectively. The continuous lines connecting the data points are shown as a guide to the eye, as $Q/e$ only can have integer values.}
\label{Fig7}
\end{figure}
%%%%%%%%%%%%%%%%%%%%%%%%%%%%%%%%%%%%%%%%%%

%%%%%%%%%
\section{Conclusions} \label{Conclusions}
%%%%%%%%%

In the present work, we extend the mean-field theory~\cite{Avni2018,Markovich2018} of mobile charge-regulated (CR) macro-ions, by taking explicitly into account the interactions between association/dissociation sites on the macro-ions. The general behavior of this system composed of macro-ions and salt ions is rather rich in complexity. Above a critical ion concentration (CIC), it becomes favorable for the macro-ions to adsorb a large number of salt ions from the bathing solution and to reach almost full occupancy of the macro-ion association/dissociation sites. As the added salt concentration surpasses the CIC value, the salt ions get further adsorbed by the macro-ions. The system now exhibits a two-phase (or even a three-phase) coexistence region, where macro-ions with low site occupation coexists with macro-ions of nearly fully saturated sites. For symmetric macro-ions, a bimodal phase in which the macro-ions are either highly positively or highly negatively charged, can persist throughout a large range of salt concentration.

Interestingly, this behavior shows a similarity with the behavior of micellar solutions close to the critical micelle concentration (CMC)~\cite{Israelachvili1992}, and even more so with a related phenomenon of micelle and vesicle formation of cationic and anionic surfactants~\cite{Marques1993, Rosa2006}. However, here the macro-ions are not formed at the critical concentration, but instead they exist regardless of the small ions, and the small ions can only change the macro-ion {\it charge state}. Moreover, since the concentrations of small ions and macro-ions are independent, we can vary either of them in order to reach the CIC point.

The coexistence of several charge states may have additional implications. In a recent and related study, the existence of several macromolecular charged states was shown to induce a pH-dependent liquid-liquid phase separation~\cite{Prost2020}, which may possibly explain the formation of membrane-less organelles in cellular biological systems, which recently became an intensely pursued research focus~\cite{Brangwynne2015, Molliex2015}. While the employed model in Ref.~{\cite{Prost2020}} is different than the one studied here, as it considers charging by protonation and deprotonation and relies on the Flory-Huggins theory, it shows that distinct charge states of macromolecules are important ingredients in a CR-driven phase separation. While in Ref.~\cite{Prost2020}, the existence of the coexisting charge states is a model assumption, here we formally derive and interpret the physics leading to it.

In order to study phase separation within our model ({\it i.e.}, as a result of the nucleation of the different charge states), one needs to go beyond the mean-field level~\cite{Netz2003}, and to introduce a consistent incorporation of the electrostatic potential fluctuations. This could be done by a one-loop expansion or, equivalently, by using the Debye-H\" uckel theory that leads to a correlation correction to the free-energy~\cite{LL1980, Levin2002}. These and related developments are left for future studies.

%%%%%%%%%%%%%%%%%%%%%%%%%%%%%%%%%%%%
\bigskip\bigskip
\noindent {\em Acknowledgements}~~
We would like to thank R. Adar, D. Frydel, T. Markovich and J. Prost for useful suggestions and correspondence.
This research has been supported by the Naomi Foundation through the Tel Aviv University
GRTF Program. R.P. would like to acknowledge the support of the {\sl 1000-Talents Program} of the Chinese Foreign Experts Bureau, and the University of the Chinese Academy of Sciences, Beijing.  D.A. acknowledges support from the Israel Science Foundation (ISF) under Grant No. 213/19 and from the NSFC-ISF Grant No. 3396/19,  and Y.A. is thankful for the support of the Clore Scholars Programme of the Clore Israel Foundation.

\appendix
%
%%%%%%%%%%%
\section{pH-dependent protonation/deprotonation mechanism} \label{appendixA}
%%%%%%%%%%%%
We apply our model to a pH-dependent protonation/deprotonation mechanism. The macro-ion surface sites are assumed to have negatively charged ionic group that are denoted by ${\rm A^-}$. They can adsorb a proton, ${\rm H^+}$, and become neutral (${\rm AH}$). If the sites are non interacting, the  protonation/deprotonation process can be described by the chemical reaction
\begin{equation} \label{AppA_eq1}
{\rm AH\rightleftharpoons A^{-}+H^{+}}.
\end{equation}
However, as we focus in the present work on cooperative processes with a coupling between protonation/deprotonation of different sites, the proper definition of the process is
\begin{equation} \label{AppA_eq2}
{\rm M}\rightleftharpoons{\rm M}^{-z}+z{\rm H^{+}},
\end{equation}
where ``M" denotes the neutral macro-ion and ``M$^{-z}$" denotes a macro-ion with $z$ negative dissociated sites. In this way, for each $z$, the reaction is association with a different energy gain and a different dissociation constant.

We note that in an acidic solution, the protons are likely to associate to water molecules, ${{\rm H_2 O}+{\rm H^+}\to{\rm H_3 O^+}}$. However, for simplicity we disregard this additional reaction, and the concentration of positive ions, $n_+$, is simply the proton concentration $[{\rm H^+}]$.

We recall that in this Appendix the adsorbed state is neutral, ${\rm AH}$, and the charging process is achieved {\it by dissociation} (unlike charging by {\it association} considered in Secs.~\ref{model} and~\ref{Results} above). Moreover, here the positive ions ${\rm H^+}$ are the only ions exchanged between the macro-ions and the solution, and the macro-ion can only become negatively charged, rather than both positive and negative as considered in the main text. Due to these differences, the internal free-energy of a single macro-ion becomes
\begin{equation} \label{AppA_eq3}
g\left(z\right)= -ez\psi-\frac{1}{\beta}\ln{N_{-} \choose z} +F_{{\rm CR}}\left(z\right)+\mu_{-}z.
\end{equation}
where $F_{\rm CR}(z)$ is the free-energy of the dissociated (rather than the associated) state. Following the same derivation as in Sec.~\ref{model_a}, we obtain
${p_{\rm b}(z)=\mathcal{A}\exp[{-\beta g_0\left(z\right)}]}$, with
\begin{equation} \label{AppA_eq4}
\begin{split}\beta g_{0}\left(z\right)= & -\ln{N \choose z}+\beta F_{{\rm CR}}\left(z\right)+z\ln\left(n_{+}^{{\rm b}}\lambda^{3}\right).
\end{split}
\end{equation}
Note that due to the dissociation charging mechamism, $p_{\rm b}(z)$ is proportional to $(n_{+}^{\rm b})^{-z}$, whereas in an association charging mechamism it is proportional to $(n_{+}^{\rm b})^z$. In other words, as the concentration of free protons is larger, the macro-ion sites are more likely to be in a neutral ${\rm AH}$ state.

Within the quadratic approximation we employed, $F_{{\rm CR}}$ becomes
\begin{equation} \label{AppA_eq5}
\beta F_{{\rm CR}}=\alpha z+\frac{\chi}{2N}z^{2}
\end{equation}
where $\alpha$ and $\chi$ are phenomenological parameters.

Next, we relate the model parameters to the solution pH, and the acid dissociation constant in Eq.~(\ref{AppA_eq2}), pK$_a (z)$, that depends on $z$ as for each $z$ the reaction may be different. The pH and pK$_a (z)$ are defined by,
\begin{equation} \label{AppA_eq6}
\begin{split}
\text{pH}&=-\log_{10}\left[{\rm H}^{+}\right]\\
\text{pK}_{a}(z)&=-\log_{10}\frac{\left[{\rm H}^{+}\right]^{z}\left[{\rm M}^{-z}\right]}{\left[{\rm M}\right]}
\end{split}
\end{equation}
and within our formalism, they become
\begin{equation} \label{AppA_eq7}
\begin{split}
\text{pH} & =-\log_{10}n_{+}^{{\rm b}}\\
\text{pK}_{a}(z)&=-\log_{10}\left[{N \choose z}\lambda^{-3z}{\rm e}^{-\alpha z-\frac{\chi}{2N}z^{2}}\right].
\end{split}
\end{equation}

\section{The non-symmetric model} \label{appendixB}
We present the corresponding analysis of Secs.~\ref{model_a} and~\ref{model_b} for a general model of macro-ions (not necessarily symmetric between positive and negative sites). Note that to recover the symmetric case, which was described in detail in the main text we substitute $\chi_{+}=\chi_{-}=\chi_1/2$, $\chi_{+-}=\chi_2/2$ and  $N=N_{\rm T}/2$.

We denote $N_{\rm T}$ as the total number of sites, ${N_{\rm T}=N_++N_-}$. Expanded to quadratic order, $F_{\rm CR}$ has the form
\begin{equation} \label{AppB_eq1}
\beta F_{{\rm CR}}=\alpha_{+}z_{+}+\alpha_{-}z_{-}+\frac{1}{2}\frac{\chi_{+}}{N_{\rm T}}z_{+}^{2}+\frac{1}{2}\frac{\chi_{-}}{N_{\rm T}}z_{-}^{2}+\frac{\chi_{+-}}{N_{\rm T}}z_{+}z_{-}
\end{equation}
where the linear terms account for the independent free-energy gain from adsorption of a positive charge and a negative charge (first and second terms, respectively), and the quadratic terms represent the free-energy changes due to short-range interactions between positive, negative and oppositely charged neighboring adsorption sites (3rd, 4th and 5th terms, respectively).

In the $N_{\rm T}\gg 1$ limit, we define as before $\tilde{g}_0\left(\phi_{+},\phi_{-}\right)=g_0\left(z_{+},z_{-}\right)/N$ and use the Stirling's formula to obtain,
\begin{equation} \label{AppB_eq2}
\begin{split}\beta\tilde{g}_{0} & =\zeta\phi_{+}\ln\phi_{+}+\zeta\left(1-\phi_{+}\right)\ln\left(1-\phi_{+}\right)\\
 & +\left(1-\zeta\right)\phi_{-}\ln\phi_{-}+\left(1-\zeta\right)\left(1-\phi_{-}\right)\ln\left(1-\phi_{-}\right)\\
 & -\zeta\phi_{+}\left[\ln\left(n_{+}^{{\rm b}}\lambda^{3}\right)-\alpha_{+}\right]\\
 & -\left(1-\zeta\right)\phi_{-}\left[\ln\left(n_{-}^{{\rm b}}\lambda^{3}\right)-\alpha_{-}\right]\\
 & +\frac{1}{2}\chi_{+}\zeta^{2}\phi_{+}^{2}+\frac{1}{2}\chi_{-}\left(1-\zeta\right)^{2}\phi_{-}^{2}\\
 & +\zeta\left(1-\zeta\right)\chi_{+-}\phi_{+}\phi_{-}.
\end{split}\end{equation}
where $\zeta$ is the fraction of sites that can be positively charged, satisfying the relations ${N_+=N_{\rm T} \zeta}$ and ${N_-=N_{\rm T}\left(1-\zeta\right)}$. By minimizing $\tilde{g}_{0}$, the average values of $\phi_{\pm}$ in the bulk are obtained as follows,
\begin{equation} \label{AppB_eq3}
\begin{split}
\frac{\phi_{+}}{1-\phi_{+}} & =n_{+}^{{\rm b}}\lambda^{3}{\rm e}^{-\alpha_{+}-\zeta\chi_{+}\phi_{+}-\left(1-\zeta\right)\chi_{+-}\phi_{-}}\\
\frac{\phi_{-}}{1-\phi_{-}} & =n_{-}^{{\rm b}}\lambda^{3}{\rm e}^{-\alpha_{-}-\left(1-\zeta\right)\chi_{-}\phi_{-}-\zeta\chi_{+-}\phi_{+}}.
\end{split}
\end{equation}
If ${\chi_{+-}=0}$, {\it i.e.}, no interaction between the different types of sites, the equations reduce to two Langmuir-Frumkin-Davies adsorption isotherms,
\begin{equation} \label{AppB_eq4}
\begin{split}\phi_{+} & =\frac{1}{1+\left(n_{+}^{{\rm b}}\lambda^{3}\right)^{-1}{\rm e}^{\alpha_{+}+\zeta\chi_{+}\phi_{+}}}\\
\phi_{-} & =\frac{1}{1+\left(n_{-}^{{\rm b}}\lambda^{3}\right)^{-1}{\rm e}^{\alpha_{-}+\left(1-\zeta\right)\chi_{-}\phi_{-}}}.
\end{split}
\end{equation}
A phase diagram, similar to the one described in Fig.~{\ref{Fig2}}, can be obtained from Eqs.~(\ref{AppB_eq2})-(\ref{AppB_eq3}). However, unlike the 3D phase diagram in the symmetric case, here the phase space has six dimensions and can be parametrized by: $\left(\zeta,\,\ln\left(n_{+}^{{\rm b}}\lambda^{3}\right)-\alpha_{+},\,\ln\left(n_{-}^{{\rm b}}\lambda^{3}\right)-\alpha_{-},\,\chi_{+},\,\chi_{-}\,,\chi_{+-}\right)$.

%%%%%%%%%%%%%%%%%%%%%%%%%%%%%%%%%%%

%%%%%%%%%%%%%%
\end{document}